\magnification=\magstep1
\input psfig.sty
\font\bigbfont=cmbx10 scaled\magstep1
\font\bigifont=cmti10 scaled\magstep1
\font\bigrfont=cmr10 scaled\magstep1
\vsize = 23.5 truecm
\hsize = 15.5 truecm
\hoffset = .2truein
\baselineskip = 14 truept
\overfullrule = 0pt
\parskip = 3 truept
\def\frac#1#2{{#1\over#2}}

%
\def\Sz{S^z}
\font\sf=cmss10
\def\prob{\hbox{\sf p}}
\def\cite#1{{\ignorespaces\lbrack#1\rbrack}}
\def\bibitem#1{\parindent=8mm\item{\hbox to 6 mm{\cite{#1}\hfill}}}
\def\etal#1{#1}
\def\DR{1}
\def\EKT{2}
\def\KFE{3}
\def\Elbio{4}
\def\Hida{5}
\def\Okamoto{6}
\def\TNKone{7}
\def\TotsukaOne{8}
\def\AOY{9}
\def\Kuramoto{10}
\def\MSBMY{11}
\def\HMT{12}
\def\YaSa{13}
\def\LaMaD{14}
\def\CHPprl{15}
\def\CHPprb{16}
\def\tandon{17}
\def\WFMK{18}
\def\COAIQ{19}
\def\GrCha{20}
\def\Haldane{21}
\def\CPR{22}
\def\BoFi{23}
\def\PaBo{24}
\def\SaTa{25}
\def\KiOk{26}
\def\BWRZZHD{27}
\def\AAIAG{28}
\def\SwWi{29}
\def\SLW{30}
\def\ITHUOTKN{31}
\def\OkKi{32}
\def\TOHTK{33}
\def\HoLa{34}
\def\ChHiNa{35}
\def\CaGy{36}
\def\poly{37}
\def\WFMKtwo{38}
\def\CHS{39}
\def\DRZ{40}
\def\CaGyTwo{41}
\def\STKTUOTMG{42}
\def\zigzag{43}
\def\CaGyThree{44}
\def\Kolezhuk{45}
\def\FrSo{46}
\def\CdMHPS{47}
\def\CdMGPP{48}
\def\Mu{49}
\def\TotsukaTwo{50}
\def\Mila{51}
\def\CJYFHBLHP{52}
\def\FuZh{53}
\def\WeHa{54}
\def\KOK{55}
\def\ToHa{56}
\def\SGMK{57}
\def\GFAMAK{58}
\def\GMK{59}
\def\UsSu{60}
\def\TNKtwo{61}
\def\SBUMH{62}
\def\White{63}
\def\HOA{64}
\def\NHSKNT{65}
\def\OHA{66}
\def\HKS{67}
\def\MHMG{68}
\def\NTM{69}
\def\condM{1}
\def\HamXXZ{2}
\def\Hbos{3}
\def\EQsz{4}
\def\EQsp{5}
\def\perta{6}
\def\pertb{7}
\def\pHub{8}
\def\perthz{9}
\font\amsmath=msbm10

\def\Zed{\hbox{\amsmath Z}}

\topinsert
\line{Contribution to \hfill cond-mat/0010376}
\line{``XXIV INTERNATIONAL WORKSHOP \hfill October 24, 2000}
\leftline{ON CONDENSED MATTER THEORIES''}
\leftline{Buenos Aires, Argentina, September 12--17, 2000}
\vskip 0.9 true cm
\endinsert
\centerline{\bigbfont Magnetization Plateaux in Quasi-One-Dimensional}
\vskip 6 truept
\centerline{\bigbfont Strongly
Correlated Electron Systems: A Brief Review}
\vskip 20 truept
\centerline{\bigifont D.C.\ Cabra$^{1,2}$, M.D.\ Grynberg$^{1}$,
A.\ Honecker$^{3}$, P.\ Pujol$^{4}$}
\vskip 8 truept
\centerline{\bigrfont $^{1}$Departamento de F\'{\i}sica, Universidad Nacional de la
Plata,}

\centerline{\bigrfont C.C.\  67, (1900) La Plata, Argentina}
\centerline{\bigrfont $^{2}$Facultad de Ingenier\'{\i}a, Universidad Nacional
           de Lomas de Zamora,}
\centerline{\bigrfont Cno.\ de Cintura y Juan XXIII, (1832), Lomas de Zamora,
     Argentina.}
\vskip 2 truept
\centerline{\bigrfont $^{3}$Institut f\"ur Theoretische Physik, TU
Braunschweig,}
\centerline{\bigrfont Mendelssohnstr.\ 3, 38106 Braunschweig, Germany.}
\vskip 2 truept
\centerline{\bigrfont $^{4}$Laboratoire de Physique,
Groupe de Physique Th\'eorique, ENS Lyon,}
\vskip 2 truept
\centerline{\bigrfont 46 All\'ee d'Italie, 69364 Lyon C\'edex 07, France.}
\vskip 1.8 truecm

\centerline{\bf 1.~Introduction}
\vskip 12 truept
The study of quasi-one dimensional strongly correlated electron
systems (SCES's) has received a lot of attention in the past few
years. This interest was mainly triggered by the synthesis
of materials which in a wide range of temperatures can be well
modeled by a 3D system of (almost) decoupled spin chains, spin
ladders or more generally, Hubbard chains and ladders \cite{\DR}.
An important reason that has put
these 1D systems again on the scene is the appearance of the
so-called ``stripe" phases which have shown up in very different
contexts, such as high-$T_c$ cuprates \cite{\EKT}, integer quantum Hall effect
systems at high Landau levels \cite{\KFE}, materials showing colossal
magnetoresistance (manganites) \cite{\Elbio}, etc.

The present article contains a brief review of theoretical work on
magnetization plateaux in SCES's in 1D. The presentation reflects the
authors' perspective and due to lack of space, not all important
contributions to this subject can be mentioned.
The first studies on this issue were performed by
Hida \cite{\Hida} and Okamoto \cite{\Okamoto} in an attempt to describe
some organic compounds with periodic couplings. After a few other isolated
cases were studied \cite{\TNKone,\TotsukaOne}, Oshikawa and
collaborators \cite{\AOY} have undertaken the first systematic study of
this problem and, by extending the Lieb-Schultz-Mattis theorem to
systems in a magnetic field, they provided a necessary condition
for the appearance of magnetization plateaux in 1D systems. When the
magnetization $\langle M \rangle$ is normalized to saturation values
$\pm 1$, this condition for the appearance of a plateau with
magnetization $\langle M \rangle$ can be cast in the form
$$S V \left(1 - \langle M \rangle \right) \in \Zed \, .
\eqno(\condM)$$
Here $S$ is the size of the local spin and $V$ the number of spins in the unit cell
for the translation operator acting on the magnetization $\langle M \rangle$
groundstate\footnote{$^1$}{In cases where the sites carry different spins
\cite{\Kuramoto-\LaMaD}, the
combination $S V$ should be replaced by the maximal spin in the unit cell.}.
It should be noted that translational invariance can be spontaneously
broken in the groundstate and then $V$ would be larger than the unit cell of
the Hamiltonian.

Spin ladders in a magnetic field constitute a class
of systems where the full phase diagram was explored and
where it was checked when the necessary condition (\condM) becomes
also sufficient \cite{\CHPprl-\GrCha}.

A famous result on pure 1D spin chains is Haldane's
conjecture \cite{\Haldane} (see also \cite{\CPR} for a new  field
theoretical approach). It states that such chains with integer $S$ are gapful while
those with half-integer $S$ remain gapless, corresponding to the condition
(\condM) with $\langle M \rangle = 0$ and $V=1$ since
a spin gap is equivalent to an $\langle M \rangle = 0$ plateau.
This feature is indeed confirmed in seminal numerical studies of
Heisenberg chains \cite{\BoFi,\PaBo}, but no other plateaux are found
for these systems.  Other non-trivial plateaux, which for $S > 1$ would be
permitted by (\condM), arise only when the model is modified, e.g.\ by adding a
single-ion anisotropy \cite{\SaTa,\KiOk}.

The existence of real materials with trimer constituents
\cite{\BWRZZHD-\ITHUOTKN} motivated the investigation of
the magnetization process of trimerized \cite{\Hida,\Okamoto,\OkKi},
frustrated trimerized \cite{\TOHTK,\HoLa}, quadrumerized \cite{\ChHiNa}
and more generally periodically modulated $S=1/2$ Heisenberg spin chains
with period $p$ (so-called $p$-merized chains) \cite{\CaGy-\DRZ}.
Ladders consisting of both staggered and non-staggered dimerization along the
chains were also studied \cite{\CaGyTwo}.

{}From the experimental point of view, one of the most exciting materials
is NH$_4$CuCl$_3$ where just two plateaux with $\langle M \rangle = 1/4$ and
$3/4$ have been observed \cite{\STKTUOTMG}. The crystal structure of
NH$_4$CuCl$_3$ suggests to model it as a two-leg zig-zag ladder, however
with room for some modifications. While one can indeed obtain plateaux with
$\langle M \rangle = 1/4$ and $3/4$ in a zig-zag ladder with dimerized legs
\cite{\zigzag,\CaGyThree}, neither this point of view nor other proposals
\cite{\Kolezhuk} have so far resulted in a really satisfactory theoretical
description of the experimental observations. In short, if NH$_4$CuCl$_3$ is a
quasi-one-dimensional system, eq.\ (\condM) should apply with $S=1/2$
since the spin is carried by Cu$^{2+}$ ions. Then one needs $V=8$ in order
to permit plateaux with $\langle M \rangle = 1/4$ and $3/4$. However,
$V=8$ would permit also plateaux with $\langle M \rangle = 0$ and $1/2$ which
are {\it not} observed experimentally \cite{\STKTUOTMG}, but are very
pronounced e.g.\ in the dimerized zig-zag ladder \cite{\CaGyThree}.

When charge degrees of freedom are dynamical, as in Hubbard
chains, the situation turns out to be quite different and very
interesting, although the effect of a magnetic field has been studied so
far essentially only in two cases:
An integrable spin-$S$ generalization of the $t-J$ model \cite{\FrSo} and
doped Hubbard chains with periodically modulated hopping matrix elements
or on-site energies \cite{\CdMHPS}. In addition to
fully gapped situations (both a charge gap and a spin gap), situations
arise in which magnetization plateaux appear at {\it irrational}
doping-dependent values of the magnetization, in contrast
to what we have just reviewed for purely magnetic systems.
Interestingly, under such circumstances,
gapless degrees of freedom are present and consequently certain
superconducting correlations have power law behavior \cite{\CdMHPS}.

Recently, the effect of disorder on plateau systems, such as the
$p$-merized $XXZ$ chains, has also been studied. It was found that, for binary
disorder, plateaux do not disappear but are rather shifted in a precise
amount which can be predicted by means of a simple argument to be
\cite{\CdMGPP} $\langle M \rangle = 1 + \frac{2}{p} \,(\prob-1)$,
where $\prob$ is the strength of disorder.
Conversely, continuous distributions of disorder erase completely
the plateau structure. Interestingly, the effect of disorder on the
susceptibility leads to an even-odd effect similar
to what is found in $N$-leg ladders \cite{\Mu}.


\vskip 28 truept

\centerline{\bf 2.~Field theory approach }
\vskip 12 truept

The field theory description of spin chains is a useful technique for
treating problems of weakly coupled spin ladders, weak dimerization in
chains, etc. as we will illustrate below for the case of $S=1/2$.
For higher spin chains, non-Abelian bosonization has proven to be
better suited.
For spin $1/2$ chains, the
Abelian bosonization technique has proven to be very efficient. It describes
the low energy, large scale behavior of the system, and can be extended to
the case of an easy axis anisotropy. More specifically, the continuum limit
of the Hamiltonian
$$
H_{XXZ} = J \sum_{x=1}^L \left\{ \Delta \Sz_x \Sz_{x+1} + {1 \over 2}
\left(S^{+}_x S^{-}_{x+1} + S^{-}_x S^{+}_{x+1} \right)\right\}
- h \sum_{x=1}^L \Sz_{x}
\eqno(\HamXXZ)
$$
is given by the Tomonaga-Luttinger Hamiltonian
$$
H = {1 \over 2} \int dx \left( v K (\partial_x \tilde{\phi})^2
+ {v \over K} (\partial_x \phi)^2 \right) \, .
\eqno(\Hbos)
$$
The bosonic field $\phi^i$ and its dual $\tilde{\phi}^i$ are given by the
sum and difference of the lightcone components, respectively. The
constant $K=K(\langle M \rangle , \Delta)$ governs the conformal
dimensions of the bosonic vertex
operators and can be obtained exactly from the Bethe Ansatz
solution of the $XXZ$ chain (see e.g.\ \cite{\CHPprb} for a detailed
summary). One has $K=1$ for the $SU(2)$ symmetric case ($\Delta =
1$) and it is related to the radius $R$ of \cite{\CHPprb} by $K^{-1} =
2 \pi R^2$.

In terms of these fields, the spin operators read
$$\eqalignno{
S_{x}^z &= {1 \over \sqrt{2\pi}} \partial_x \phi
+ a : \cos(2 k_F x + \sqrt{2 \pi} \phi): + \frac{\langle M
\rangle}{2} \, , &(\EQsz) \cr
S_{x}^{\pm} &= (-1)^x
:{\rm e}^{\pm i\sqrt{2\pi} \tilde{\phi}}
\left(b \cos(2 k_F x + \sqrt{2 \pi} \phi) + c \right) : \, ,
&(\EQsp)
}$$
where the colons denote normal ordering with respect to the
groundstate with  magnetization $\langle M\rangle$. The Fermi
momentum $k_F$ is related to the magnetization of the chain
as $k_F = (1-\langle M \rangle )\pi/2$. An $XXZ$ anisotropy and/or the
external magnetic field modify the scaling dimensions of the physical
fields through $K$ and the commensurability properties of the spin
operators, as can be seen from (\EQsz), (\EQsp).

If one introduces periodically modulated couplings between the spins
$J_x = J$ if $x \neq np$ and $J_x = (1- \delta) J \equiv J'$ if $x = np$, this
amounts to the following perturbation for the original Hamiltonian in the
continuum limit \cite{\CaGy}
$$
H_{pert} = \lambda \int ~ dx ~ \cos(2  p k_F x + \sqrt{2 \pi}
\phi)\ ,
\eqno(\perta)
$$
where $\lambda$ is proportional to $\delta J$.

Another interesting situation is the one of spin ladders \cite{\CHPprl,\CHPprb},
where $N$ identical chains are coupled with a transversal coupling $J'$
$$
H = \sum_{a=1}^N H^a_{XXZ} + J'
    \sum_{x,a=1}^{a =N} \vec{S}^a_x \cdot \vec{S}^{a+1}_x \, .
$$
For simplicity we used here periodic boundary conditions (PBC's) along the
transverse direction. One obtains in the continuum a collection of identical
Hamiltonians like (\Hbos), with perturbation terms which couple the
fields of the different chains. After a careful renormalization group
(RG) analysis, one can show that at most one degree of freedom, given by the
combination of fields $\phi_D = \sum_a \phi_a$, remains massless.
The large scale effective action for the ladder systems is then given again
by a Hamiltonian (\Hbos) for $\phi_D$ and the perturbation term
$$
H_{pert} = \lambda \int ~ dx ~ \cos(2  N k_F x + \sqrt{2 \pi} \phi_D) \,,
\eqno(\pertb)
$$
where $k_F = (1-\langle M \rangle )\pi/2$ 
is related to the total magnetization $\langle M \rangle$.

In both cases, the key point if to identify the values of the magnetization
for which the perturbation operators (\perta) or (\pertb)
can play an important r\^ole. In fact, this operator is
commensurate at values of the magnetization given by (\condM)
with $S=1/2$ and $V = p$ or $V=N$, respectively.
If this operator turns out to be also relevant in the RG sense (this depends
on the parameters of the effective Hamiltonian (\Hbos), the model will
have a finite gap, implying a plateau in the magnetization curve. The
generalization to $p$-merization in $N$-leg ladders results in
$V = p N$, though in some cases the topology of
couplings has to be carefully analyzed \cite{\CaGyTwo}

The case of a single $p$-merized chain can be generalized to the presence of
charge carriers \cite{\CdMHPS}. Let us consider the Hubbard model,
representing interacting electrons with spin $1/2$. The Hamiltonian is
given by
$$\eqalign{
H =& - \sum_{x,\alpha} t(x) \ (c^{\dagger}_{x+1,\alpha} c_{x,\alpha}
+ H.c.) + U \sum_x c^{\dagger}_{x,\uparrow} c_{x,\uparrow}
c^{\dagger}_{x,\downarrow} c_{x,\downarrow} \cr
& + \sum_{x,\alpha} \mu (x) \ c^{\dagger}_{x,\alpha} c_{x,\alpha}
- {h \over 2} \sum_x ( c^{\dagger}_{x,\uparrow} c_{x,\uparrow} -
c^{\dagger}_{x,\downarrow} c_{x,\downarrow} ) ~,
}\eqno(\pHub)
$$
where $t(x)$ and $\mu (x)$ are taken as periodic in the variable $x$
with period $p$.
The bosonized version of this Hamiltonian can be written as a Gaussian part,
given by
$$
\sum_{i= c,s} {u_i \over 2} \int dx ~
\left[ \left( \partial_x \phi_i \right)^2 +
\left( \partial_x \theta_i \right)^2 \right] ~,
$$
where the fields labeled by $c$ and $s$ are particular combinations of
the bosonic fields dictated by the Bethe Ansatz solution of the
homogeneous model. The most relevant of several perturbation terms is given by
$$
O_{pert} =
\lambda_1 \sin [{\pi n\over 2} + p n \pi x - \sqrt{\pi}\xi \phi_c]
\cos [\sqrt{2\pi}\phi_s ] + \lambda_2 \sin[ \pi n + 2 p n \pi x
-  \sqrt{4\pi} \xi \phi_c] ~.
\eqno(\perthz)
$$
We have written the perturbation term for the case of zero magnetic field for
simplicity.
Using the same arguments of commensurability and relevance as above, we can
show that for $p n \in \Zed$, we have a charge gap. If the condition is further
constrained to $p n/2 \in \Zed$, for zero magnetic field we
have also a spin gap, implying a plateau with $\langle M \rangle = 0$. In
the case of non-zero magnetic field, if one of the conditions
$$
{p \over 2} \left(n \pm \langle M \rangle \right) \in \Zed ~,
$$
is satisfied, and the doping is kept fixed (as is natural from the point of
view of experimental
realizations of doped systems), the system has a magnetization plateau,
but still exhibits massless behavior as well, e.g.\ in the specific heat
which vanishes linearly as the temperature goes to zero.
If both conditions are simultaneously satisfied, the system is
gapped in the charge and spin sectors; this situation is in fact the
generalization to arbitrary doping of the results for $p$-merized
Heisenberg chains discussed above.

\vskip 28 truept

\centerline{\bf 3.~Strong coupling arguments}
\vskip 12 truept

The magnetization process is easy to understand if the system
decouples into clusters of $V$ sites \cite{\CHPprl,\CHPprb,\poly,\CdMHPS}.
These `strongly coupled' clusters magnetize
independently such that at zero temperature the magnetization
$\langle M \rangle$ can only take finitely many values. For a spin-system
with spin $S$ they are subject to the quantization condition
(\condM) where $V$ is directly the number of spins in one of the
decoupled clusters. Using (higher order) series expansions around
this decoupling point \cite{\CHPprl,\CHPprb,\HoLa,\poly}, the
quantization condition obtained in this simple manner can be argued to be valid
also for more general parameters.

In its simplest form, this argument explains magnetization plateaux
which arise because the unit cell of the Hamiltonian contains several spins.
However, the argument can be refined to account also for spontaneous
breaking of translational symmetry by a period of two. The transitions
between plateau states can be treated by degenerate perturbation theory,
leading to effective Hamiltonians which in many cases turn out to be effectively
an $XXZ$ chain (\HamXXZ) if only the lowest order(s) are kept
\cite{\TotsukaTwo-\WeHa,\tandon,\zigzag,\HoLa}.
If the effective $XXZ$ anisotropy turns out to be sufficiently
large ($\Delta > 1$), a gap opens and translational symmetry is
spontaneously broken. In this manner one finds a further
plateau precisely in the middle between the two values of
$\langle M \rangle$ predicted by considering just the decoupling limit
itself. This illustrates that $V$ in (\condM) should be taken from the
unit cell of the groundstate whose size is an integer multiple (in general
larger than 1) of the unit cell of the Hamiltonian.

Amusingly, first-order strong-coupling results for the magnetization
process remain exact over a finite range of parameters in certain
systems with local conservation laws \cite{\HMT,\KOK}.

\vskip 28 truept

\topinsert
\centerline{\psfig{figure=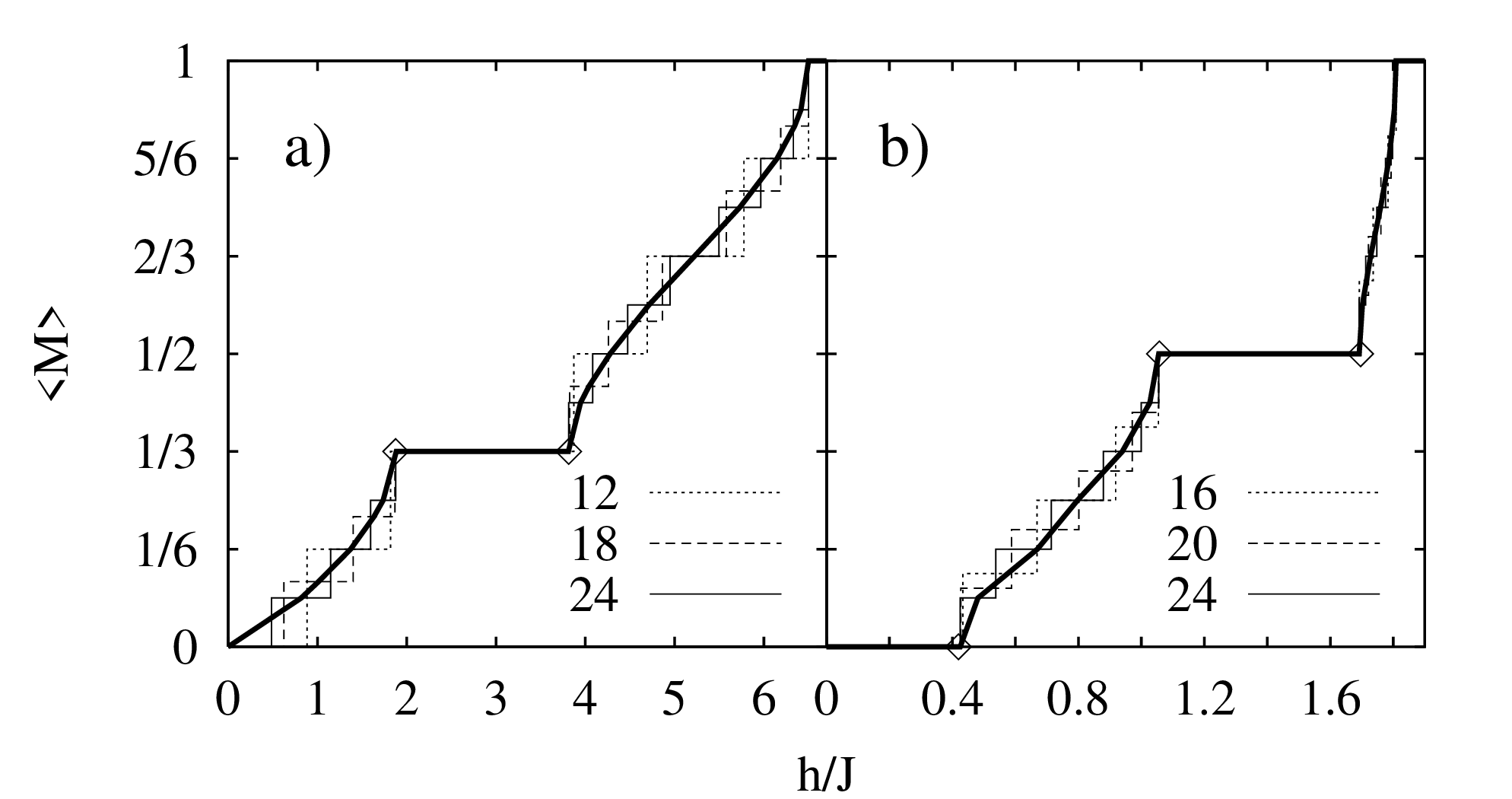,width=14.25truecm,angle=270}}
\vskip 0.4truecm
\noindent
{\bf Figure 1.\ Magnetization curves of (a) the planar 3-leg ladder
at $J' = 3J$ \cite{\CHPprl} and (b) the quadrumerized chain with $J' = J/2$
\cite{\CaGy}.
The diamonds show results of (a) fourth- \cite{\CHPprb} and (b)
second-order \cite{\poly} strong-coupling expansions for the
plateau boundaries. The bold line is an
extrapolation to the thermodynamic limit.}

\vskip 12truept
\endinsert

\centerline{\bf 4.~Numerical approaches}
\vskip 12 truept

In many cases coupling constants do not lie in any of the weak- or
strong-coupling regimes discussed before.
In such cases, accurate results can be obtained only numerically.
Since on the one hand, vector spaces
are large already for moderate system sizes and on the other hand, only a few
extreme eigenvalues are needed to compute the zero-temperature
magnetization curves, one frequently resorts to the Lanczos
method as was already done in pioneering works \cite{\BoFi,\PaBo}.
In the present situation the magnetic field couples to a conserved
quantity $S^z_{\rm tot}$. One therefore only needs the
groundstate energy $E(S^z_{\rm tot},h=0)$ at $h=0$ in each of
the magnetization subspaces with
$S^z_{\rm tot} \in \left\{0,\,1,\,...\,, S L\right\}$
(here $L$ is the total number of spins). Then one can readily obtain the energy
in a finite field $h$ through the relation
$E(S^z_{\rm tot},h) = E(S^z_{\rm tot},0) - h S^z_{\rm tot}$
and then construct the groundstate magnetization curve. Note that
the magnetization has steps on a finite system since the
possible values of the magnetization are quantized.


Among the systems whose magnetization process has been studied
by the Lanczos method, the two-leg zig-zag ladder (which is
equivalent to the $J_1$-$J_2$ chain) has a particularly long
tradition \cite{\ToHa-\TNKtwo}. However, plateaux with
$\langle M \rangle \ne 0$ are observed only for more than two
legs \cite{\zigzag,\WFMKtwo} or if the coupling constants are
modulated \cite{\CaGyTwo,\CaGyThree,\TotsukaTwo,\TNKtwo,\SBUMH}.

Fig.\ 1a) shows the magnetization curve of an $S=1/2$
3-leg ladder with open boundary conditions
along the rungs \cite{\CHPprl}. As expected, there is a clear plateau with
$\langle M \rangle = 1/3$ and the fourth-order strong coupling
series for the boundaries of the plateau \cite{\CHPprb} are
in good agreement with the Lanczos results.
The effect of $p$-merization is illustrated in Fig.\ 1b) which
shows the magnetization curve of a quadrumerized $S=1/2$ Heisenberg
chain \cite{\CaGy}. Here one observes two clear plateaux at
$\langle M \rangle = 0$ and $1/2$, as expected. The series
for their boundaries \cite{\poly} are in good agreement with the
numerical data even if they are only of second order.
Moreover, in the latter case bosonization predicts the plateaux
to open with power laws in $\delta$ which can indeed be verified
numerically \cite{\CaGy}.


\topinsert
\centerline{\psfig{figure=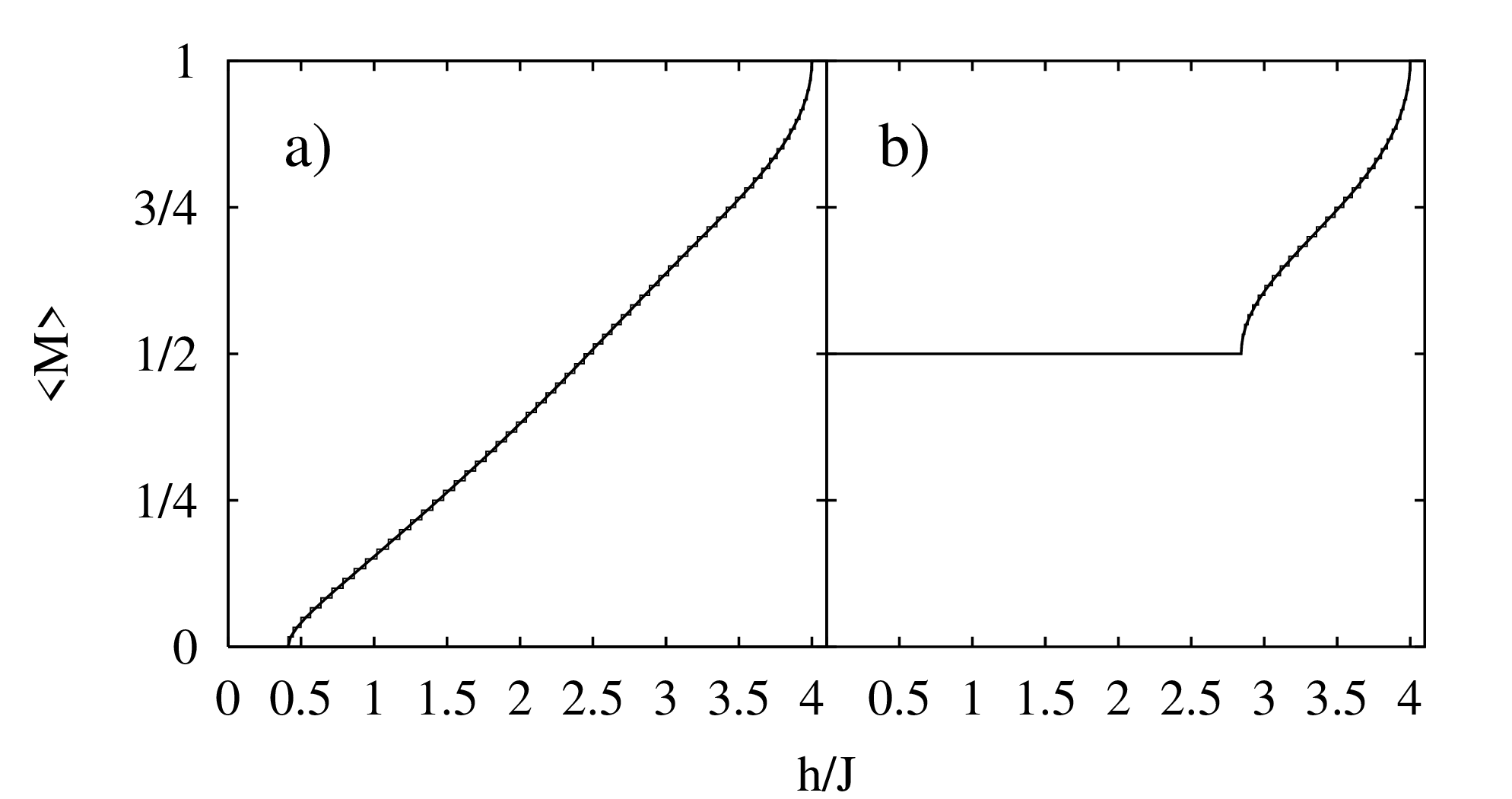,width=14.25truecm,angle=270}}
\noindent
{\bf Figure 2.\ Magnetization curves of (a) the $S=1$ Heisenberg chain
and (b) the $S=3/2$-$1/2$ ferrimagnetic Heisenberg chain.
The steps show the curves obtained for a finite system with
$L=60$ sites by DMRG \cite{\HMT} while the smooth curves denote an
extrapolation to the thermodynamic limit.}

\vskip 12truept
\endinsert

%

%

%

An altenative method is DMRG \cite{\White} which allows to study larger
systems. The number of applications of this method or variants thereof to
the magnetization process is increasing steadily
\cite{\MSBMY,\HMT,%
\tandon-\COAIQ,
\TOHTK,%
\WFMKtwo,%
\SBUMH,%
\HOA-\HKS}. 
The only technical complication is that the infinite system algorithm
does not give good approximations to incommensurate structures and
therefore the finite system algorithm should be used if one wants to compute
the complete magnetization curve. Fig.\ 2 illustrates this method with
two examples \cite{\HMT}: (a) The magnetization curve of the
$S=1$ Heisenberg chain exhibits only an $\langle M \rangle = 0$
plateau, corresponding to the Haldane gap \cite{\Haldane},
and (b) the magnetization curve of the $S=3/2$-$1/2$ ferrimagnetic
Heisenberg chain which exhibits a spontaneous magnetization
and thus a plateau with $\langle M \rangle = 1/2$.

\topinsert
\centerline{\psfig{figure=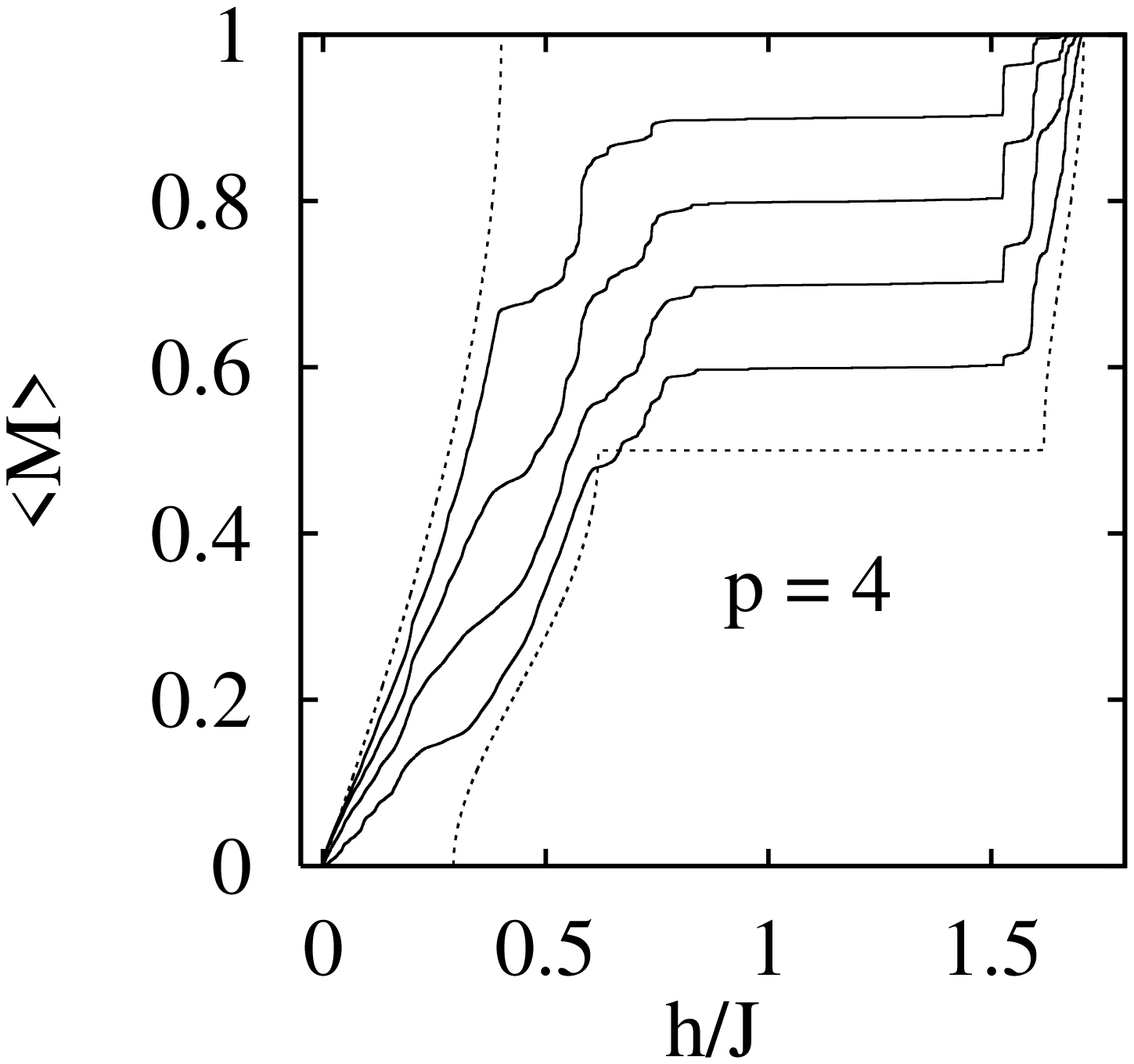,width=7.6truecm,angle=00}}
\noindent
{\bf Figure 3.\ Magnetization curves of quadrumerized $XX$ chains
with $5 \times 10^4$ sites in disordered binary backgrounds of strength
$\prob = 0.2,\, 0.4,\, 0.6, \,0.8$ (solid lines in ascending order)
\cite{\CdMGPP}.
The left and rightmost dotted lines denote respectively the pure uniform
and pure modulated cases $\prob = 1$ and $\prob = 0$.}

\vskip 12truept
\endinsert

Long $XX$ chains in a transversal field can be treated by the Jordan-Wigner
transformation (see e.g.\ \cite{\DRZ}). This is particularly useful
for systems with quenched disorder \cite{\CdMGPP}.
Fig.\ 3 illustrates the effect of a disordered distribution of
couplings with a $p$-periodically modulated background on the magnetic
behavior. Here, we show the magnetization curves of a quadrumerized $XX$
chain under different disorder strengths $\prob$ of a binary distribution
$P(J_i) =  \prob \delta(J_i-J') + (1-\prob) \delta(J_i-J_0 - \gamma_i J)$,
where $\gamma_i = \gamma \,, (-\gamma)$ if $i = p n\,,$ ($ i \ne p n$).
One observes the appearance of
a rather nontrivial phenomenon, namely, the shift of the magnetization
values (not necessarily rational), for which certain plateaux
emerge, as compared to the pure system. One can show by means of a simple
real space decimation procedure \cite{\CdMGPP}
that these new plateaux appear at $\langle M \rangle = 1 + 2 (\prob-1)/p$
for $XXZ$ chains which in the special case of $XX$ chains is consistent with
the numerical results. Continuous distributions for the disorder
wipe out completely the plateau structure.

\vskip 28 truept

%

\centerline{\bf 5.~Experimental realizations}
\vskip 12 truept

Two concrete plateau substances have already been mentioned earlier:
Clear plateaux with $\langle M \rangle = 1/4$ and $3/4$ have been observed
in NH$_4$CuCl$_3$ \cite{\STKTUOTMG}, but in this case a thorough theoretical
understanding is still lacking. Conversely, a plateau with
$\langle M \rangle = 1/3$ is theoretically expected \cite{\TOHTK,\HoLa} in
Cu$_3$Cl$_6$(H$_2$O)$_2\cdot$2H$_8$C$_4$SO$_2$,
but magnetization experiments \cite{\ITHUOTKN} would have to be extended to
slightly higher fields to confirm its presence.
One example where experimental observations do actually confirm theoretical
predictions \cite{\TNKone,\TotsukaOne}
are the plateaux with $\langle M \rangle = 1/2$ and $0$
which have been observed in the bond-alternating $S=1$ chain compound
$[$Ni$_{2}$(Medpt)$_{2}$($\mu$-ox)($\mu$-N$_{3}$)$]$ClO$_{4}\cdot$0.5H$_{2}$O
\cite{\NHSKNT}.

$N$-leg spin ladders are realized in modifications of high-$T_c$ materials
\cite{\DR}. However, magnetization plateaux are predicted for these systems
at fields of (several) thousand Tesla which is clearly outside the
presently accessible experimental range. One alternative is provided
by organic compounds and indeed magnetization experiments have been
successfully performed on the two-leg ladder material
Cu$_2$(C$_{5}$H$_{12}$N$_{2}$)$_2$Cl$_4$ (see e.g.\ \cite{\CJYFHBLHP}).
Vanadium oxides may in general provide another alternative and e.g.\ some PBC
3-leg ladder is suspected in the recently synthesized compound Na$_2$V$_3$O$_7$
\cite{\MHMG}. However, there is already one experimental observation which
can be interpreted in terms of a 3-leg ladder: A plateau-like feature
at $\langle M \rangle = 1/3$ has been observed in CsCuCl$_3$
for a magnetic field perpendicular to the crystal axis \cite{\NTM}. Although
CsCuCl$_3$ is strictly speaking three-dimensional, this plateau can be
understood in the approximation of a weakly anisotropic PBC 3-leg
ladder with ferromagnetic rungs \cite{\HKS}.

In general, it is intriguing that doping can push magnetization plateaux
into the low-field region at least for $p$-merized Hubbard chains
\cite{\CdMHPS}. This points towards the possibility of observing
magnetization plateaux in doped systems whose parent compounds would
exhibit plateaux only at inaccessibly large fields.

\vskip 15 true pt

{\it Acknowledgements:}
%
%
D.C.C.\ and M.D.G.\ acknowledge financial support from CONICET and Fundaci\'on
Antorchas.

\vskip 28 truept

\centerline{\bf References}
\vskip 12 truept

\tolerance=10000

\bibitem{\DR} E.~Dagotto, T.M.~Rice, Science~{\bf 271}, 618 (1996);
              T.M.~Rice, Z.~Phys.~{\bf B103}, 165 (1997);
              E.\ Dagotto, Rep.\ Prog.\ Phys.\ {\bf 62}, 1525 (1999).
\bibitem{\EKT} V.J.\ Emery\etal{, S.A.\ Kivelson, J.M.\ Tranquada},
              Proc.\ Natl.\ Acad.\ Sci.\ USA {\bf 96}, 8814 (1999).
\bibitem{\KFE} E.\ Fradkin, S.A.\ Kivelson, V.J.\ Emery, Nature
               {\bf 393}, 550 (1998).
\bibitem{\Elbio} T.\ Hotta, Y.\ Takada, H.\ Koizumi, E.\ Dagotto,
              Phys.\ Rev.\ Lett.\ {\bf 84}, 2477 (2000).
\bibitem{\Hida} K.\ Hida, J.\ Phys.\ Soc.\ Jpn.\ {\bf 63}, 2359 (1994).
\bibitem{\Okamoto} K.\ Okamoto, Solid State Commun.\ {\bf 98}, 245 (1996).
\bibitem{\TNKone} T.\ Tonegawa, T.\ Nakao, M.\ Kaburagi,
              J.\ Phys.\ Soc.\ Jpn.\ {\bf 65}, 3317 (1996).
\bibitem{\TotsukaOne} K.\ Totsuka, Phys.\ Lett.\ {\bf A228}, 103 (1997).
\bibitem{\AOY} M.\ Oshikawa, M.\ Yamanaka, I.\ Affleck,
              Phys.\ Rev.\ Lett.\ {\bf 78}, 1984 (1997).
\bibitem{\Kuramoto} T.\ Kuramoto, J.\ Phys.\ Soc.\ Jpn.\ {\bf 67}, (1998)
              1762.
\bibitem{\MSBMY} K.\ Maisinger\etal{, U.\ Schollw\"ock, S.\ Brehmer, H.-J.\
              Mikeska, S.\ Yamamoto}, Phys.\ Rev.\ {\bf B58}, (1998) R5908.
\bibitem{\HMT} A.\ Honecker, F.\ Mila, M.\ Troyer,
              Eur.\ Phys.\ J.\ {\bf B15}, 227 (2000).
\bibitem{\YaSa} S.\ Yamamoto, T.\ Sakai, Phys.\ Rev.\ {\bf B62}, 3795 (2000).
\bibitem{\LaMaD} A.\ Langari, M.A.\ Mart\'{\i}n-Delgado, mpi-pks/0002003,
              Phys.\ Rev.\ {\bf B}, in press.
\bibitem{\CHPprl} D.C.\ Cabra, A.\ Honecker, P.\ Pujol,
              Phys.\ Rev.\ Lett.\ {\bf 79}, 5126 (1997).
\bibitem{\CHPprb} D.C.\ Cabra, A.\ Honecker, P.\ Pujol, Phys.\ Rev.\ {\bf B58},
              6241 (1998).
\bibitem{\tandon} K.\ Tandon\etal{, S.\ Lal, S.K.\ Pati, S.\ Ramasesha, D.\
              Sen}, Phys.\ Rev.\ {\bf B59}, 396 (1999).
\bibitem{\WFMK} R.\ Wie{\ss}ner\etal{, A.\ Fledderjohann, K.-H.\ M\"utter,
              M.\ Karbach}, Phys.\ Rev.\ {\bf B60}, 6545 (1999).
\bibitem{\COAIQ} R.\ Citro\etal{, E.\ Orignac, N.\ Andrei, C.\ Itoi, S.\ Qin},
              J.\ Phys.: Condensed Matter {\bf 12}, 3041 (2000).
\bibitem{\GrCha} D.\ Green, C.\ Chamon, cond-mat/0004292,
              Phys.\ Rev.\ Lett., in press.
\bibitem{\Haldane} F.D.M.\ Haldane, Phys.\ Rev.\ Lett.\ {\bf 50}, 1153 (1983);
               Phys.\ Lett.\ {\bf A93}, 464 (1983).
\bibitem{\CPR} D.C.\ Cabra, P.\ Pujol, C.\ von Reichenbach,
              Phys.\ Rev.\ {\bf B58}, 65 (1998).
\bibitem{\BoFi} J.C.\ Bonner, M.E.\ Fisher, Phys.\ Rev.\ {\bf 135}, A640 (1964).
\bibitem{\PaBo} J.B.\ Parkinson, J.C.\ Bonner, Phys.\ Rev.\ {\bf B32},
              4703 (1985).
\bibitem{\SaTa} T.\ Sakai, M.\ Takahashi, Phys.\ Rev.\ {\bf B57}, R3201 (1998).
\bibitem{\KiOk} A.\ Kitazawa, K.\ Okamoto, Phys.\ Rev.\ {\bf B62}, 940 (2000).
\bibitem{\BWRZZHD} M.R.\ Bond\etal{, R.D.\ Willett, R.S.\ Rubins, P.\ Zhou,
              C.E.\ Zaspel, S.L.\ Hutton, J.E.\ Drumheller},
              Phys.\ Rev.\ {\bf B42}, 10280 (1990).
\bibitem{\AAIAG} Y.\ Ajiro\etal{, T.\ Asano, T.\ Inami, H.\ Aruga-Katori,
              T.\ Goto}, J.\ Phys.\ Soc.\ Jpn.\ {\bf 63}, 859 (1994).
\bibitem{\SwWi} D.D.\ Swank, R.D.\ Willett,
              Inorganica Chimica Acta {\bf 8}, 143 (1974).
\bibitem{\SLW} D.D.\ Swank\etal{, C.P.\ Landee, R.D.\ Willett},
              J.\ Magnetism and Magnetic Materials {\bf 15}, 319 (1980).
\bibitem{\ITHUOTKN} M.\ Ishii\etal{, H.\ Tanaka, M.\ Hori, H.\ Uekusa, Y.\
              Ohashi, K.\ Tatani, Y.\ Narumi, K.\ Kindo},
              J.\ Phys.\ Soc.\ Jpn. {\bf 69}, 340 (2000).
\bibitem{\OkKi} K.\ Okamoto, A.\ Kitazawa,
              J.\ Phys.\ A: Math.\ Gen.\ {\bf 32}, 4601 (1999).
\bibitem{\TOHTK} T.\ Tonegawa\etal{, K.\ Okamoto, T.\ Hikihara, Y.\ Takahashi,
              M.\ Kaburagi}, J.\ Phys.\ Soc.\ Jpn.\ {\bf 69} Suppl.\ A, 332
              (2000).
\bibitem{\HoLa} A.\ Honecker, A.\ L\"auchli, cond-mat/0005398.
\bibitem{\ChHiNa} W.\ Chen, K.\ Hida, H.\ Nakano, J.\ Phys.\ Soc.\ Jpn.\
               {\bf 68}, 625 (1999).
\bibitem{\CaGy} D.C.\ Cabra, M.D.\ Grynberg, Phys. Rev. {\bf B59}, 119 (1999).
\bibitem{\poly} A.\ Honecker, Phys.\ Rev.\ {\bf B59}, 6790 (1999).
\bibitem{\WFMKtwo} R.M.\ Wie{\ss}ner\etal{, A.\ Fledderjohann, K.-H.\ M\"utter,
              M.\ Karbach}, Eur.\ Phys.\ J.\ {\bf B15}, 475 (2000).
\bibitem{\CHS} W.\ Chen, K.\ Hida, B.C.\ Sanctuary, cond-mat/0009481.
\bibitem{\DRZ} O.\ Derzhko, J.\ Richter, O.\ Zaburannyi, cond-mat/0010120.
\bibitem{\CaGyTwo} D.C.\ Cabra, M.D.\ Grynberg, Phys.\ Rev.\ Lett.\ {\bf 82},
               1768 (1999).
\bibitem{\STKTUOTMG} W.\ Shiramura\etal{, K.\ Takatsu, B.\ Kurniawan, H.\
              Tanaka, H.\ Uekusa, Y.\ Ohashi, K.\ Takizawa, H.\ Mitamura, T.\
              Goto}, J.\ Phys.\ Soc.\ Jpn.\ {\bf 67}, 1548 (1998).
\bibitem{\zigzag} D.C.\ Cabra, A.\ Honecker, P.\ Pujol,
              Eur.\ Phys.\ J.\ {\bf B13}, 55 (2000).
\bibitem{\CaGyThree} D.C.\ Cabra, M.D.\ Grynberg, Phys.\ Rev.\ {\bf B62},
              337 (2000).
\bibitem{\Kolezhuk} A.K.\ Kolezhuk, Phys.\ Rev.\ {\bf B59}, 4181 (1999).
\bibitem{\FrSo} H.\ Frahm, C.\ Sobiella, Phys.\ Rev.\ Lett.\ {\bf 83}, 5579
              (1999).
\bibitem{\CdMHPS} D.C.\ Cabra\etal{, A.\ De Martino, A.\ Honecker, P.\ Pujol,
              P.\ Simon}, Phys.\ Lett.\ {\bf A268}, 418 (2000);
              cond-mat/0008333.
\bibitem{\CdMGPP} D.C.\ Cabra\etal{, A.\ De Martino, M.D.\ Grynberg,
 S.\ Peysson, P.\ Pujol}, cond-mat/0007205,
 to appear in Phys. Rev. Lett.
\bibitem{\Mu} P.W.\ Brouwer, C.\ Mudry, A.\ Furusaki,
              Phys.\ Rev.\ Lett.\ {\bf 84}, 2913 (2000).
\bibitem{\TotsukaTwo} K.\ Totsuka, Phys.\ Rev.\ {\bf B57}, 3454 (1998);
              Eur.\ Phys.\ J.\ {\bf B5}, 705 (1998).
\bibitem{\Mila} F.\ Mila, Eur.\ Phys.\ J.\ {\bf B6}, 201 (1998).
\bibitem{\CJYFHBLHP} G.\ Chaboussant\etal{, M.-H.\ Julien, Y.\ Fagot-Revurat,
              M.\ Hanson, C.\ Berthier, L.P.\ L\'evy, M.\ Horvati\'c,
              O.\ Piovesana}, Eur.\ Phys.\ J.\ {\bf B6}, 167 (1998).
\bibitem{\FuZh} A.\ Furusaki, S.C.\ Zhang,
              Phys.\ Rev.\ {\bf B60}, 1175 (1999).
\bibitem{\WeHa} S.\ Wessel, S.\ Haas, Eur.\ Phys.\ J.\ {\bf B16}, 393 (2000);
              Phys.\ Rev.\ {\bf B62}, 316 (2000).
\bibitem{\KOK} A.\ Koga, K.\ Okunishi, N.\ Kawakami, Phys.\ Rev.\ {\bf B62},
              5558 (2000).
\bibitem{\ToHa} T.\ Tonegawa, I.\ Harada, J.\ Phys.\ Soc.\ Jpn.\ {\bf 56},
              2153 (1987);
              Physica {\bf B155}, 379 (1989);
              J.\ Phys.\ Soc.\ Jpn.\ {\bf 58}, 2902 (1989).
\bibitem{\SGMK} M.\ Schmidt\etal{, C.\ Gerhardt, K.-H.\ M\"utter, M.\ Karbach},
              J.\ Phys.: Condensed Matter {\bf 8}, 553 (1996).
\bibitem{\GFAMAK} C.\ Gerhardt\etal{, A.\ Fledderjohann, E.\ Aysal, K.-H.\
              M\"utter, J.F.\ Audet, H.\ Kr\"oger},
              J.\ Phys.: Condensed Matter {\bf 9}, 3435 (1997).
\bibitem{\GMK} C.\ Gerhardt, K.-H.\ M\"utter, H.\ Kr\"oger, Phys.\ Rev.\ {\bf
              B57}, 11504 (1998).
\bibitem{\UsSu} M.\ Usami, S.\ Suga, Phys.\ Lett.\ {\bf A240}, 85 (1998).
\bibitem{\TNKtwo} T.\ Tonegawa, T.\ Nishida, M.\ Kaburagi,
              Physica {\bf B246-247}, 368 (1998).
\bibitem{\SBUMH} F.\ Sch\"onfeld\etal{, G.\ Bouzerar, G.S.\ Uhrig, E.\
              M\"uller-Hartmann,} Eur.\ Phys.\ J.\ {\bf B5}, 521 (1998).
\bibitem{\White} S.R.\ White, Phys.\ Rev.\ Lett.\ {\bf 69}, (1992) 2863;
              Phys.\ Rev.\ {\bf B48}, (1993) 10345.
\bibitem{\HOA} Y.\ Hieida, K.\ Okunishi, Y.\ Akutsu, Phys.\ Lett.\ {\bf A233},
              464 (1997).
\bibitem{\NHSKNT} Y.\ Narumi\etal{, M.\ Hagiwara, R.\ Sato, K.\ Kindo, H.\
              Nakano, M.\ Takahashi}, Physica {\bf B246-247}, 509 (1998).
\bibitem{\OHA} K.\ Okunishi, Y.\ Hieida, Y.\ Akutsu, Phys.\ Rev.\ {\bf B59},
              6806 (1999).
\bibitem{\HKS} A.\ Honecker, M.\ Kaulke, K.D.\ Schotte, Eur.\ Phys.\ J.\ {\bf
              B15}, 423 (2000) 423.
\bibitem{\MHMG} P.\ Millet, J.Y.\ Henry, F.\ Mila, J.\ Galy, J.\ Solid State
              Chem.\ {\bf 147}, 676 (1999).
\bibitem{\NTM} H.\ Nojiri\etal{, Y.\ Tokunaga, M.\ Motokawa},
              Journal de Physique {\bf 49}, Suppl.\ C8, 1459 (1988).

\end